\documentclass[12pt]{iopart}
\usepackage[dvips]{graphicx}
\usepackage{amssymb}

\def\beq{\begin{equation}}
\def\eeq{\end{equation}}
\def\beqn{\begin{eqnarray}}
\def\eeqn{\end{eqnarray}}

\begin{document}

\title{A No-go theorem for Poincar\'e-invariant networks}

\author{Sabine Hossenfelder}

\address{Nordita, 
KTH Royal Institute of Technology and Stockholm University\\
Roslagstullsbacken 23, SE-106 91 Stockholm, Sweden}
\ead{hossi@nordita.org}
%\vspace{10pt}
%\begin{indented}
%\item[]February 2014
%\end{indented}

\begin{abstract}
I explain why there are no Poincar\'e-invariant networks with a locally
finite distribution of nodes in Minkowski-spacetime of any dimension. 
\end{abstract}

% Uncomment for PACS numbers
%\pacs{00.00, 20.00, 42.10}
%
% Uncomment for keywords
%\vspace{2pc}
%\noindent{\it Keywords}: XXXXXX, YYYYYYYY, ZZZZZZZZZ
%
% Uncomment for Submitted to journal title message
%\submitto{\JPA}
%
% Uncomment if a separate title page is required
%\maketitle
% 
% For two-column output uncomment the next line and choose [10pt] rather than [12pt] in the \documentclass declaration
%\ioptwocol
%

\section{Introduction}

Discretization is a traditional cure for divergences. For this reason many scenarios have been proposed that try to deal with the non-renormalizability of quantum
gravity by replacing the continuous
space-time of general relativity with a discrete background \cite{Loll:1998aj}. 
However, problems with recovering
Lorentz-invariance in the suitable limits will
inevitably occur in any discretization that makes use of networks or triangulations. Therefore, all 
approaches relying on a
space-time discretization have to address whether this Lorentz-invariance
violation leads to conflicts with existing observations.

The following argument is probably intuitively clear to most people working 
in the field, but to my best knowledge it has not been put into writing before. 
These notes are to benefit newcomers and those not very familiar with the
literature.

\section{Invariance versus Covariance}

Before we go in media res, we first have to clarify what we mean with Poincar\'e-invariant because there is a common misunderstanding in the terminology. Any
object that is specified by quantities which do not change under any Poincar\'e-transformations is Poincar\'e-invariant. This is the case for scalars and four-volumes
(properly weighted), and combinations thereof. 

A tensor of rank higher than zero is not invariant, it is covariant. This means it transforms
from one coordinate system to the other by a well-defined procedure. 
A similar statement is true for tensor-densities. Any tensor known in one coordinate-system can be made
into a Poincar\'e-covariant tensor by just defining it in all other coordinate
systems as the transformation of the quantity in one coordinate system. Thus
the question whether or not a tensor is Poincar\'e-invariant is only meaningful
if the expression of the tensor is known in several coordinate systems.

Not any object that has four entries is a covariant vector. Take for example
the diagonal entries of the stress-energy tensor. These do {\it not} transform
like a vector. Likewise, not every $4\times4$ matrix is a 2nd rank tensor. This
is to say that the correct definition of physically meaningful quantities bears relevance,
and in fact one of Einstein's main achievement was to identify suitable combinations
of physical quantities so that they would form tensors.

The expression of a tensor of rank larger than zero in one coordinate system can be obtained uniquely
from its expression in another coordinate system, but the tensor entries themselves
will in general be different and so these objects are not invariants. The difference between
invariance and covariance is relevant if one is concerned with the question
of Lorentz-invariance violation. We say Lorentz-invariance (and thereby Poincar\'e-invariance) is 
violated if one can construct a preferred space-time
slicing, or a preferred class of observers respectively. 

What we mean here with `preferred slicing'
is any kind of procedure that allows arbitrary observers to determine a particular time-like
vector-field $(\partial_t)^\alpha$ from possible measurements they can make. This vector field is covariant and one can
construct a corresponding space-like slice $\Sigma_t$ at each point in time which is locally orthogonal
to the vector field. Using the vector field and the corresponding slices constitutes then a specific $3+1$ split
of space-time. Typically, the laws giving rise to the measurements which the observers made
will be particularly simple in the coordinates in which the vector field is just $(-1,0,0,0)$ which
is why this frame is said to be `preferred'.

It must be emphasized though that the existence of preferred frames does not mean
that there are now quantities that do no longer transform under Lorentz-transformations from
one inertial frame to the other\footnote{Since Lorentz-transformations are just a special case of coordinate transformations
in Minkowski-space, the attempt to do physics with laws of nature that do not
transform like tensors is extremely problematic and will in the following not be
considered.}. 

\begin{quote}
Definition: {\it Any structure that allows to construct a preferred slicing of space-time or
a preferred direction in space is
{\bf Lorentz-invariance violating}. A structure that does not allow such construction
 is {\bf Lorentz-invariant}.} 
\end{quote}

We often use preferred frames in our calculations. On Earth for example,
a frame in rest to the Earth's surface is preferred for calculations in the laborary settings, and similarly a frame in rest with the cosmic
microwave background is a preferred frame for cosmological observables. These frames are not
worrisome because they arise from the dynamics and distribution of matter within space, but are
not a fundamental ingredient of our theories independent of the matter. 
We know that on short distances the interactions of matter are described by the standard model and respect local Lorentz-invariance
in all higher order corrections. The problem we are concerned with here is a fundamental vector field that is
not merely an emergent field assigned to dynamical matter but is a property of space-time itself. The effects of
such a vector field will not normally go away on short distances. Instead, these fundamental
vector fields can come to make unduly large contributions to standard model processes. 

Violations of Lorentz-invariance by the presence of a preferred frame or a preferred
direction are highly
constrained by existing experiments because one would expect such a vector field
to couple to the Standard Model and make itself noticeable in particle
interactions. Moreover, one expects that
in an effective limit any Lorentz-invariance violating fundamental space-time
structure gives rise to such a preferred vector field. Or if it doesn't, this at least
necessitates a strong argument as for why a fundamental Lorentz-invariance violation
should vanish in the effective limit. It was pointed out already in \cite{Collins:2004bp} that
Lorentz-invariance violation induced by quantum gravity causes a fine-tuning problem,
and the worry that even small Lorentz-invariance
violations can lead to large effects was further fleshed out in \cite{Gambini:2011nx,Gambini:2014kba}.

The strength of existing constraints
depends of course on exactly which particles the field couples to. Constraints
are particularly strong if the vector field that defines the preferred frame couples
to electrons or photons. In the case where
the field couples only gravitationally, constraints are comparably weak. The 
purpose of this paper isn't to elaborate on experimental constraints on Lorentz-invariance
violation and its parameterization, but just to explain that it has to be paid attention to in any
approach that replaces space-time with a discrete structure. For more details
and constraints on Lorentz-invariance violation, please refer to \cite{Mattingly:2005re}.

To conclude this section I should be fair and admit that much of the confusion
around this comes about because in the literature
the distinction between Lorentz-invariance and Lorentz-covariance isn't always
properly made. I too have almost certainly sometimes sloppily said `invariance' when what 
I really meant was `covariance'. The phrase `observer-independence' can for
this reason also be ambiguous because it might either mean that all observers really
see the same (invariant) or that they can transform their observations into each
other's by a well-defined procedure (covariant). However, the exact physical meaning
often reveals itself from the given context.

\section{Theorem}

Any regular space-time lattice, that is a network with nodes connected by links, violates Poincar\'e-invariance. This is so to begin with because 
the lattice is not
translationally invariant under continuous displacements, but in a sense this rather uninteresting.
One could argue that as long as one looks at large enough scales, this discretization of
translations would be unobservable. The more severe problem is
that Lorentz-boosts make the very notion of what is a `large' scale meaningless and
so these violations cannot easily be limited --
if one just boosts sufficiently much, even a wavelength similar to the Planck-length can be redshifted
to a kilometer. A regular lattice is defined by its lattice vectors that 
do not differ from one node to the
next. These vectors are covariant, but
not invariant. Their (scalar) space-time interval is invariant, but their spatial components 
can become arbitrarily large or arbitrarily small. One can use them to construct a preferred
frame for example by defining it to be the frame in which the spatial spacing is just
exactly a Planck length. Pictorially speaking, the lattice isn't invariant because its
spacing is stretched and squeezed under Lorentz-boosts. The same holds true for
non-periodic quasi-crystals because the tiles have typical widths.

We will assume here that the discrete space-time network approximates
continuous space-time and that it is possible to create a distance measure
on that emergent space-time which corresponds to the metric. We can
then assign relative distances to the nodes of the network. The 
Poincar\'e transformations act on these distances. 

It has been known for some while however that there are Poincar\'e-invariant
random distributions of points in Minkowski-space that are locally finite \cite{Bombelli:2006nm}. Locally
finite means that there are finitely many points in any finite space-time volume,
which is what one needs for discretization to be meaningful. These invariant distributions can be explicitly constructed by a Poincar\'e-process. 
They are invariant in the stochastical sense, so that averaged over a large number
of repetitions (or a large sample of volumes respectively) they are invariant, even
though any single point distribution on its own is not. 

We will not need the explicit expression of this Poisson-process
for the following, but its mere existence raises the question whether not one
can use this distribution of points to make it into a network that is also both
Poincar\'e-invariant (on the average) and locally finite. With a locally finite
network we will mean the following:

\begin{quote}

Definition: {\it A {\bf locally finite network} is a network in which the number of nodes is finite
in any finite space-time volume and the number of links pinching the surface of
each volume is finite.}
\end{quote}

Having clarified all necessary expressions, we can then make precise the statement
we want to prove:
\begin{quote}
Theorem: {\it There exist no Poincar\'e-invariant networks in Minkowski-space that are locally finite.} 
\end{quote}

\subsection{Proof in 1+1 dimensional Minkowski-space}

We will lead this proof by contradiction. For simplicity we first do this in 1+1 dimensional
Minkowski-space. The generalization to higher dimensions is then straight-forward.

\begin{figure}

{
\includegraphics[height=8cm]{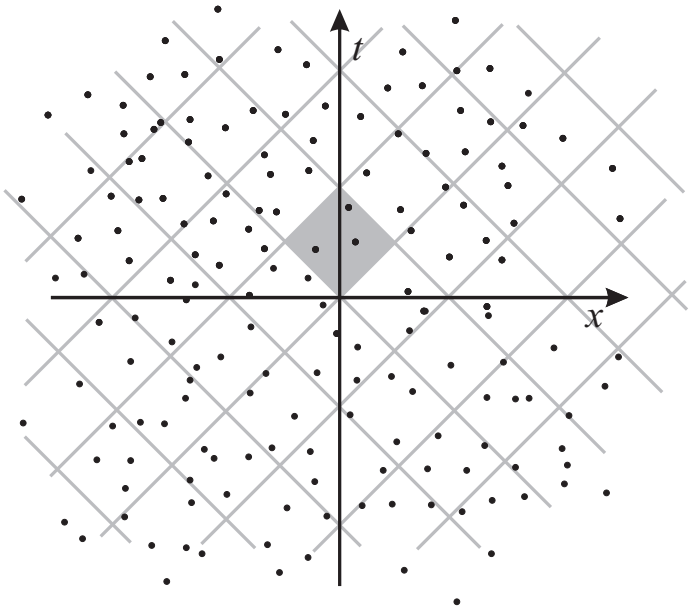} } 

\caption{First we divide up space-time into tiles of equal volume. Dots are nodes of the
network. This is a sketch -- the shown distribution is not actually Poincar\'e-invariant on the
average.  \label{fig1}}
\end{figure}

Suppose there exists a locally finite Poincar\'e-invariant network. This network must have
a finite amount of nodes in any finite volume. We thus take the 1+1 dimensional Minkowski-space
and tile it up into causal diamonds of some fixed spatial and temporal extension, see Figure \ref{fig1}. This tiling
is not Lorentz-invariant, but this does not matter in the following. Relevant is only that the
tiles all have the same volume and thus, due to homogeneity must, on the average, contain the
same finite number of nodes.

Next we pick one of these tiles (marked grey in Figure \ref{fig1}) and one node in this tile and ask what can be
the network neighbors of this node so that the network connections are invariant on the average. Well,
first the node might be entirely unconnected, which is of course Lorentz-invariant, but if all
nodes were like this it would be a pretty uninteresting type of network, so we will not further consider this
possibility. Consider then the node has at least one neighbor, where is it?

The node and
its neighbor define a space-time distance $(\Delta t, \Delta x)$. The space-time
interval defined by the proper length $\alpha := - \Delta t^2 + \Delta x^2$ is Lorentz-invariant
and we need not worry about it. The spatial and temporal component however depend on
the reference frame and could be used to construct a preferred frame, unless they are
uniformly distributed over all values with the same $\alpha$. This means that the
probability for the position of the neighboring point must be uniformly distributed on the hyperbolae defined by $\alpha$. 

%..........................................................................
%..........................................................................

\begin{figure}
{\includegraphics[height=8cm]{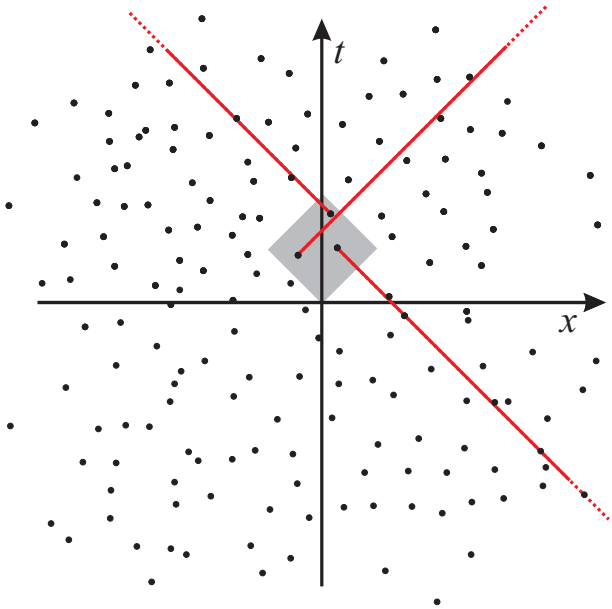}  }

\caption{Next we try to find out where a node in the center tile might have its
neighbors. Since the probability for the endpoints must be uniformly distributed
on each hyperbola at constant proper length from the origin, the neighbors are
almost certainly located at an arbitrarily large spatial distance, ie arbitrarily close
to the lightcone. \label{fig2}}
\end{figure}

The Lorentz-group is non-compact and so the boost parameters cover an infinite
range, corresponding to an infinite length of the hyperbolae. Uniform distributions
on non-compact support are awkward -- the mathematician would say they
are ill-defined and would go on to now introduce all kinds of epsilons and deltas
to handle the issue. The physicist who is more comfortable with infinities
just concludes that the neighboring node is with probability one in the infinite spatial
and temporal distance, see Figure \ref{fig2}. 

For a uniform probability distribution
of points on an infinite interval, the probability to find the point in any finite interval
is zero. If we pick for example the top right quadrant, this means the neighboring 
node is to be found at $\lim_{\Delta x \to \infty} (\sqrt{\Delta x^2 - \alpha},\Delta x)$.
Another way to say this is that the only vectors invariant under a Lorentz-transformation
are $(0, 0)$ and $(\pm \infty, \pm \infty)$, understood as a limiting case. The nodes in the 
center tile might have several more neighbors for which the same
conclusion applies, but taking these into account is not necessary for
the following.

%..........................................................................
%..........................................................................

We could now select whether we want the node to be only in the forward or
in the backward lightcone, depending on whether we have a directed or
undirected network, but this isn't so relevant for the following. So let us just
assume that with probability $1/2$ the neighbor is either in the future
or the past and with probability $1/2$ either left or right. 
 
At this point one may already question what the physical meaning is of such a
network that aligns its links along lightcones, but it seems too early to give up.
So let us then ask next now many links go through the volume that we had
picked. 

\begin{figure}

\includegraphics[width=8cm]{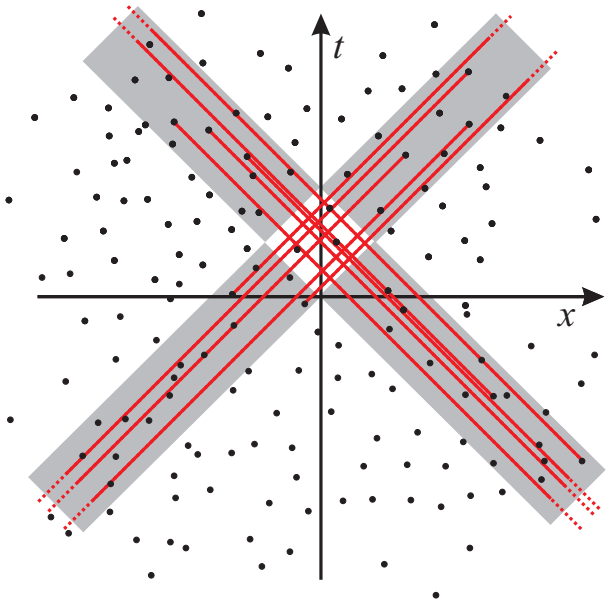}  

\caption{Finally we count all the links that go through the center node. Since
there are infinitely many points with links arbitrarily close to the lightcone, there
are infinitely many links passing through the center tile, and thus through every tile. \label{fig3}}
\end{figure}

Because all the links are arbitrarily close to the lightcones, we must
count all the points in the volume-tiles that can send light-signals to or receive
light signals from the volume we considered (see Figure \ref{fig3}). Each
of these points has a probability of $1/4$ to have a link that goes through
the center volume. Since there are infinitely many tiles, there are infinitely
many  links going through the surface of the
center volume and thereby, due to homogeneity, through every other volume.
So we conclude that the network isn't locally finite $\Box$.

%..........................................................................
%..........................................................................

\subsection{Minkowski-space of higher dimension}

In the case of more than one spatial dimension the argument of the previous
subsection has to modified so as to take into account that the probability for
the endpoint is now distributed over all spatial dimensions, and also that
the volume from which links can go through the center tile is larger. These
two factors cancel each other: The probability that a link from one of the
volume tiles in the light-like past or future of the center tile goes through
the center tile drops with $1/D^2$, where $D$ is the spatial distance (in
the chosen coordinate system), while the number of tiles from which a
link can reach the center tile increases with $1/D^2$. This means if one
integrates over all links the result is still infinite. (And in this infinite limit
it is independent of the reference frame. If one cuts off the integral at
any finite slice, then the result will depend on how that slice was chosen.)

If this sounds familiar, it is probably because the same argument leads to Olber's paradox,
according to which the night sky should be infinitely bright in an infinite
universe that is homogeneously and isotropically populated with stars. 
In this case it is the luminosity of the star that falls with $1/D^2$,
while the number of stars at a given distance increases with $D^2$. 

\subsection{Partitions by hypersurfaces}

The problem of infinitely dense networks carries over to partitions of space-time
by two-dimensional hypersurfaces, like for example triangulations. These cannot be made Poincar\'e-invariant
while being locally finite for the same reason as networks, because these
surfaces, like one-dimensional lines but unlike points and space-time volumes, are merely
covariant but not invariant under coordinate transformations. The easiest way to see
this is to take the intersection of such a partition with an arbitrary timelike slice of
Minkowski-space. This will result in a network on the slice which should also
be Poincar\'e-invariant, and thus the above considerations apply.

\subsection{Links that are not straight}

It should be added for completion that the above conclusion cannot be
avoided if the links of the network are not straight lines. If the network (the nodes and links,
for example described by a matrix) is
thought to be fundamental, such a notion doesn't make much sense, but
one may think of this as a generalized kind of network, for example
like the string-network considered in \cite{Afshordi:2014cia}. 

If the connections are continuous but not straight lines, a volume enclosing
them must have
a finite maximal spatial width\footnote{And a non-continuous connection seems an oxymoron.}. For this width to be Lorentz-invariant it thus
again can either be zero (getting us back to the straight line) or infinite (which 
cannot be because otherwise the curve wouldn't connect the links). Thus, considering
curves rather than straight lines just makes the problem worse because
it introduces more quantities whose Lorentz-invariance has to be assured. 

\section{Discussion} 
\label{disc}

Now that we have seen just what the problem is with making networks
Poincar\'e-invariant let us discuss some ways to circumvent these problems.

First, one can avoid links and surfaces and only deal with points and volumes.
This is exactly the idea of the causal set approach \cite{Bombelli:1987aa}. The causal set is a
set of points with an order relation; it is not a network and thus does not
suffer from the above mentioned problem. In the causal set approach, neighbors
of points are only constructed to the end of allowing propagation of particles
and fields, giving rise to `chains' on which particles move \cite{Ilie:2005qg}. This propagation 
however depends on initial values (a momentum for example) that 
singles out a preferred direction. One does not expect the
propagation of a particle with some initial momentum to be invariant, but
merely covariant, and consequently
there is no problem with picking neighbors to construct these chains. 

It was
argued in \cite{Moore:1988zz} that the causal set cannot be considered nodes
of a network on reasoning similar to the one put forward here but for causal
sets specifically. In \cite{Moore:1988zz}  it was also not emphasized that the
homogeneity of the distribution is essential to be able to draw the conclusion.

In \cite{Niayesh} it was pointed out that if one does not consider the full Lorentz-group but only
a subgroup whose boost parameter is discretized, then at least in 1+1 dimensional
Minkowski-space there exist distributions of points that are regular and
invariant under this subgroup. These point distributions were named `Lorentzian
lattices' in \cite{Niayesh}, but note that they do not have connections between the points, ie
do not have lattice vectors. If
one tries to add connections between the points of these distributions, then
one sees easily that that these define a preferred frame which is not invariant under the subgroup.

The second possibility to deal with the absence of Poincar\'e invariant
networks is to make sure that all
arising observables that are covariant rather than invariant depend on a 
physical process that gives rise to a preferred frame locally. 
This consideration was used in \cite{own1,own2}, in which the
distribution of space-time defects is pointlike and thus can be (on the average)
Poincar\'e-invariant, while the scattering on the defects depends on the 
momentum of the incoming particle and is thus not invariant, but covariant.
The network constructed in \cite{Boguna:2013fof} is locally
finite, but observer-dependent.

The third possibility is to avoid the infinitely dense network by making
space-time non-homogeneous or time-dependent, or both. After all, our universe
has a preferred frame, which could be taken for example as the restframe
of the cosmic microwave background. The expansion of the universe providing
us with the preferred frame also means that stars haven't lived forever
and why, last time you looked, the night sky wasn't infinitely bright.
 
The challenge with this possibility of just using a preferred frame
is to make sure that the violations of Poincar\'e-invariance are unobservable
at scales where space-time should to good precision be locally flat. 
Since this is the main point for writing this note, it is worth reflecting
on it for a bit. If it was possible to construct Poincar\'e-invariant
networks in Minkowski-space, then there would be no reason to
doubt that a network in a Friedmann-Robertson-Walker space would
reinstall Lorentz-invariance to good precision when space-time
curvature can be neglected. But since we have seen now it is not possible 
to construct such networks, it must be checked in each case just what
happens to the preferred frame in local interactions. The worry is, as
mentioned previously, that particles will locally couple to the vector field defining the
frame.

It should be added that one may discard homogeneity if one is dealing
with momentum-space rather than space-time. One can thus in principle
discretize momentum space while maintaining Lorentz-invariance. However,
since momenta are the generators of translations, the both spaces are
related to each other and it must be ensured that no contradictions
arise.

Fourth, one may try to keep a network that has a Poincar\'e-invariant
distribution of nodes, but have connections that are not Lorentz-invariant,
such as suggested in \cite{Afshordi:2014cia}. Such
networks are easy to create based on the causal set's Poisson-sprinkling.
We can for example put a tile of arbitrary but fixed volume around each node and connect
that node with all points in that tile. This breaks Lorentz-invariance
because the tiling wasn't invariant -- the resulting network will have 
a typical spatial length
of connections that depends on the restframe. In this case too one
is then however left with the question whether the Lorentz-invariance violation gives rise to any observables.

Fifth, one might start out with a network that violates Poincar\'e-invariance
but eventually take a limit in which it becomes infinitely dense and so one recovers
Poincar\'e-invariance in this limit. This is similar to taking the limit of
zero lattice spacing. While this has much promise to reinstall local
Lorentz-invariance when quantities are suitably defined, it raises the
question whether in this limit one will still enjoy the benefits of introducing
a space-time discretization to begin with, or if not one will just get back
the problems with perturbatively quantized gravity. 

Sixth, one could try to just accept that the network isn't locally
finite. As with the previously listed probability, in this case the question
is whether that has any benefits over just using continuous space-time
to begin with.

Finally, I want to emphasize that this does not mean that network-based
discretizations of space-time cannot work as a basis of quantum gravity 
because they necessarily lead to Lorentz-invariance violations that conflict with
observations. That all locally finite networks must violate Poincar\'e-invariance
merely means that one has reason to suspect that
they will give rise to an observable preferred frame, and one must
find a convincing argument why the inevitable Lorentz-invariance violation
of the underlying structure is not noticeable by us. Just referring to the
existence of Poincar\'e-invariant point-distributions is not sufficient to
make the problem go away. 

\section{Summary}

I have demonstrated here why there cannot be Poincar\'e-invariant networks
in Minkowski-space that do not have an infinitely dense coverage by links, and
have discussed the challenges this poses for discrete approaches to quantum gravity.

\section*{Acknowledgements}

I thank Niayesh Afshordi and Stefan Scherer for helpful feedback. I also want to
thank those people who commented on my blog that prompted me to write this note.

\end{document}